\documentclass[final,5p,times,twocolumn]{elsarticle}

\usepackage{amssymb}
\usepackage{amsthm}
\usepackage{amsmath}
\usepackage{array}
\usepackage{graphicx}
\usepackage{theorem}
\usepackage{dcolumn}
\usepackage{bm}
\usepackage{setspace}
\usepackage{stfloats}
\usepackage{subcaption}
\usepackage{caption}
\usepackage{textgreek}
\usepackage{xcolor}

\biboptions{sort&compress}

\usepackage{xtab}
\xentrystretch{-0.1}

\usepackage{lineno}

\journal{Int. J. Heat Mass Transfer}

\begin{document}

\begin{frontmatter}



\title{Heat and mass transfer during a sudden loss of vacuum in a liquid helium cooled tube - Part IV: Freeze range}


\author[label1,label2]{Shiran Bao}
\author[label2,label3]{Yuan Tang}
\author[label2,label3]{Qutadah Rababah}
\author[label2,label3]{Wei Guo\corref{c1}}
\address[label1]{Institute of Refrigeration and Cryogenics, Zhejiang University, Hangzhou 310027, China}
\address[label2]{National High Magnetic Field Laboratory, 1800 East Paul Dirac Drive. Tallahassee, Florida 32310, USA}
\address[label3]{Mechanical Engineering Department, FAMU-FSU College of Engineering, Florida State University, Tallahassee, Florida 32310, USA}
\cortext[c1]{Corresponding author: wguo@magnet.fsu.edu}

\begin{abstract}
In a series of papers (i.e., Part I--III), we presented our systematic study on nitrogen gas propagation in an evacuated copper tube cooled by liquid helium. This research aims to gain insights into the coupled heat and mass transfer processes involved in a catastrophic vacuum loss in superconducting particle accelerator beamline tubes. Our experimental measurements on nitrogen gas propagation in both normal liquid helium (He I) and superfluid helium (He II) cooled tube revealed a nearly exponential deceleration of the gas propagation. A theoretical model that accounts for the gas dynamics, heat transfer, and condensation was also developed, and various key experimental observations were nicely reproduced in our model simulations. A particularly interesting phenomenon uncovered in our study is that the gas propagation appears to nearly stop beyond a certain distance from the location where condensation starts. We refer to this distance as the freeze range. In this paper, we present our detailed analysis of the freeze range at various inlet mass fluxes and tube diameters. We show that the results can be well described by a simple correlation. The physical mechanism underlying this useful correlation is explained. Knowing the freeze range may allow accelerator engineers to develop protocols for controlling frost-layer contamination in the beamline tubes, which should be of practical importance.
\end{abstract}


\end{frontmatter}


\section{Introduction}
Many modern particle accelerators utilize superconducting radio-frequency (SRF) cavities cooled by liquid helium (LHe) to accelerate particles~\cite{Padamsee-2009-RFSuperCon}. These SRF cavities are housed inside interconnected cryomodules and form a long LHe-cooled vacuum tube, i.e., the beamline tube~\cite{pagani-2005-SRF05}. A vacuum failure in the beamline tube can be catastrophic because the air leaking into the beamline tube can propagate down the tube and condense on the tube inner surface, causing violent boiling of the LHe and dangerous pressure build-up in the cryomodule~\cite{wiseman1994loss,seidel2002failure,ady2014-CERN,boeckmann2008-ICE, dalesandro2012-AIP}. Understanding how fast the condensing gas propagates and the associated heat deposition is of great practical importance for the design of beamline safety components.

This motivation has triggered some focused research over the past decade~\cite{dalesandro2023}. For instance, early studies at accelerator labs showed that the propagation speed of the condensing gas was significantly lower than that in a room-temperature vacuum tube~\cite{boeckmann2008-ICE,Ady-2014-5th-IPAC-Proc.}. Later studies by Dhuley and Van Sciver revealed a nearly exponential deceleration of the gas-front propagation in LHe-cooled tubes~\cite{dhuley-2016_Int.J.HeatMassTransf.,dhuley-2016_Int.Cryog.Eng.Conf.2}. They attributed this strong deceleration to the condensation of the gas on the tube inner surface, but a quantitative explanation was lacking. In a series of recent papers~\cite{garceau-2019_Cryogenics,garceau-2019_InternationalJournalofHeatandMassTransfer,bao-2020_InternationalJournalofHeatandMassTransfer,garceau-2021_InternationalJournalofHeatandMassTransfer}, we reported our more systematic studies on nitrogen gas propagation in an evacuated copper tube immersed in both normal liquid helium (He I) and superfluid helium (He II). Besides measuring the gas-front propagation velocity in a setup under well controlled conditions~\cite{garceau-2019_Cryogenics}, we also developed a theoretical model that accounts for the gas dynamics, condensation, and heat transfer. Various key observations in both the He I runs and the He II runs were nicely reproduced in our model simulations~\cite{bao-2020_InternationalJournalofHeatandMassTransfer,garceau-2021_InternationalJournalofHeatandMassTransfer}, which greatly improved our understanding of the complex coupled heat and mass transfer processes involved in loss of vacuum events.

A particularly interesting phenomenon observed in our past studies is that at a given inlet mass flow rate, the gas propagation appears to nearly stop beyond a certain distance from the location where the condensation starts. We denote this distance as the freeze range. Knowing the freeze range may allow accelerator engineers to develop protocols for better controlling frost-layer contamination in the beamline tube. This paper presents our systematic numerical study of this freeze range. In Sec.~\ref{sec:model}, we briefly outline our theoretical model and its validation against our experimental observations. In Sec.~\ref{sec:result}, we present the calculated freeze range under various inlet mass flux and tube diameters. We show that the results can be well described by an simple correlation. In Sec.~\ref{sec:discussion}, we discuss the physical mechanism underlying the simple correlation. A brief summary is included in Sec.~\ref{sec:sum}.


\section{\label{sec:model}Theoretical model}
\begin{figure*}[!tb]
	\centering
	\includegraphics[width=1.7\columnwidth]{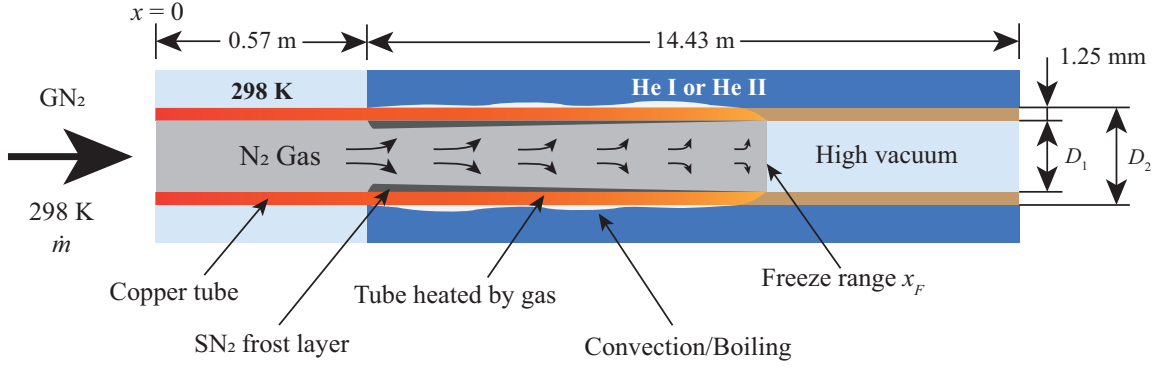}
	\caption{A schematic diagram showing the propagation and deposition of GN$_2$ in the LHe-cooled copper tube.
		\label{fig:fig1}}
\end{figure*}

We consider nitrogen gas (GN$_{2}$) propagation in an evacuated copper tube cooled by LHe as shown schematically in Fig.~\ref{fig:fig1}. At the inlet, the gas is supplied at a constant mass flow rate $\dot m$ with a fixed temperature $T=298$ K. The tube-wall thickness is set to 1.25 mm, matching that in our experiments. As detailed in our previous work~\cite{bao-2020_InternationalJournalofHeatandMassTransfer, garceau-2021_InternationalJournalofHeatandMassTransfer}, we adopt a 1D model to describe the gas dynamics and condensation in the cold section of the tube (i.e., $x>0.57$~m). This model is based on the conservation equations of the GN$_{2}$'s mass, momentum, and energy, as listed below:
\begin{equation}
	\frac{\partial\rho_g}{\partial t}+\frac{\partial}{\partial x}(\rho_g v) = -\frac{4}{D_1} \dot{m}_{c},
	\label{eq:modConsMass}
\end{equation}
\begin{equation}
	\frac{\partial}{\partial t}(\rho_g v) + \frac{\partial}{\partial x}(\rho_g v^2) = -\frac{\partial 	P}{\partial x} - \frac{4}{D_1}\dot{m}_{c} v,
	\label{eq:modConsMome}
\end{equation}
\begin{equation}
	\begin{aligned}
		\frac{\partial}{\partial t}\left[\rho_g\left(\varepsilon_g+ \frac{1}{2} v^2\right)\right]+\frac{\partial}{\partial x}\left[\rho_g v \left(\varepsilon_g+\frac{1}{2} v^2 + \frac{P}{\rho_g}\right)\right]=\\
		-\frac{4}{D_1} \dot{m}_{c} \left(\varepsilon_g+\frac{1}{2}v^2 + \frac{P}{\rho_g} \right) - \frac{4}{D^2_1} Nu\cdot k_g (T_g-T_s).
		\label{eq:modConsEner}
	\end{aligned}
\end{equation}
The definitions of the parameters included in the above equations are listed in the Nomenclature table. The Nusselt number $Nu$ for the convective heat transfer between the GN$_{2}$ and the tube wall is calculated using the Sieder-Tate correlation~\cite{incropera-2007_book}. The parameter $\dot m_c$ denotes the nitrogen mass deposition rate per unit wall inner surface area, which is evaluated using the Hertz-Knudsen relation~\cite{collier1994convective}:
\begin{equation}
	\dot{m}_{c} = \sqrt{ \frac{M_g}{2\pi R}}\left({\Gamma\sigma_c}\frac{P}{\sqrt{T_g}}-\sigma_e\frac{P_s}{\sqrt{T_s}} \right).
	\label{eq:mc}
\end{equation}
where $P_s$ is the saturated vapor pressure at the surface temperature $T_s$ of the nitrogen frost layer on the tube inner surface. The empirical condensation coefficient $\sigma_c$ and the evaporation coefficient $\sigma_e$ are set to be 0.95 in our simulations, since their values are typically about the same and close to unity for a very cold surface~\cite{persad-2016_Chem.Rev.}. The Schrage parameter $\Gamma$ is evaluated based on the GN$_{2}$ temperature $T_g$ and the mass deposition rate $\dot m_c$ as detailed in Ref.~\cite{garceau-2021_InternationalJournalofHeatandMassTransfer}. The frost-layer surface temperature $T_s$, which is required in the above equations, can be determined by evaluating the heat transfer through the frost layer~\cite{bao-2020_InternationalJournalofHeatandMassTransfer, garceau-2021_InternationalJournalofHeatandMassTransfer}:
\begin{equation}
	\rho_{SN} C_{SN} \delta \frac{\partial T_c}{\partial t} = \dot{m}_c(\frac{v^2}{2} + {\hat{h}}_g - {\hat{h}}_s) + \frac{1}{D_1}Nu\cdot k_g(T_g-T_s) - q_{con}.
	\label{eq:frost}
\end{equation}
where $T_c=(T_w+T_s)/2$ denotes the frost layer center temperature, and $q_{con}=k_{SN}(T_s-T_w)/\delta$ is the heat flux conducted through the frost layer thickness $\delta$, whose change rate is given by $\dot{\delta}=\dot{m}_c/\rho_{SN}$\,\cite{stephan1987-Physical-chemical-ref,scott1976-Physics_report}. The SN$_{2}$ properties are taken from Refs.~\cite{cook-1976_Cryogenicsa, scott1976-Physics_report}. The variation of the copper tube wall temperature $T_w$ is described by:
\begin{equation}
	\rho_w C_{w} \frac{D^2_2 - D^2_1}{4D_1} \frac{\partial T_w}{\partial t} = q_{con} - q_He \frac{D_2}{D_1}+\frac{D^2_2 - D^2_1}{4D_1} k_w \frac{\partial^2 T_w}{\partial x^2}.
	\label{eq:wall}
\end{equation}
\begin{figure}[!tb]
	\centering
	\includegraphics[width=0.9\columnwidth]{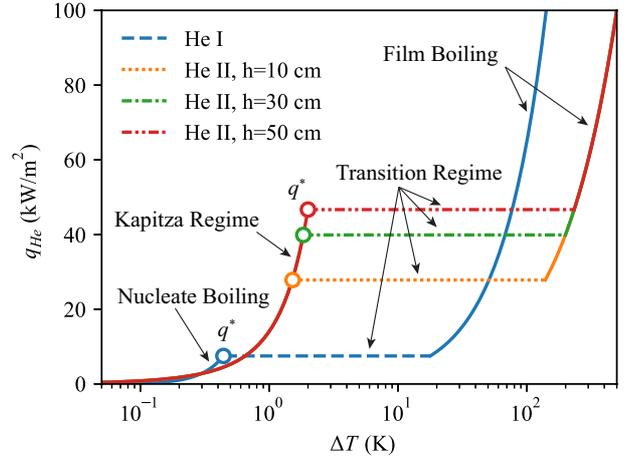}
	\caption{Representative correlation curves for the heat flux $q_He$ from the tube outer surface to the LHe bath as a function of the temperature difference $\Delta T=T_w-T_b$ between the tube wall and bath. $T_b$ is set to 4.2~K for the He I case and 1.9~K for the He II case.
	\label{fig:fig2}}
\end{figure}
\begin{figure*}[!tb]
	\centering
	\includegraphics[width=2\columnwidth]{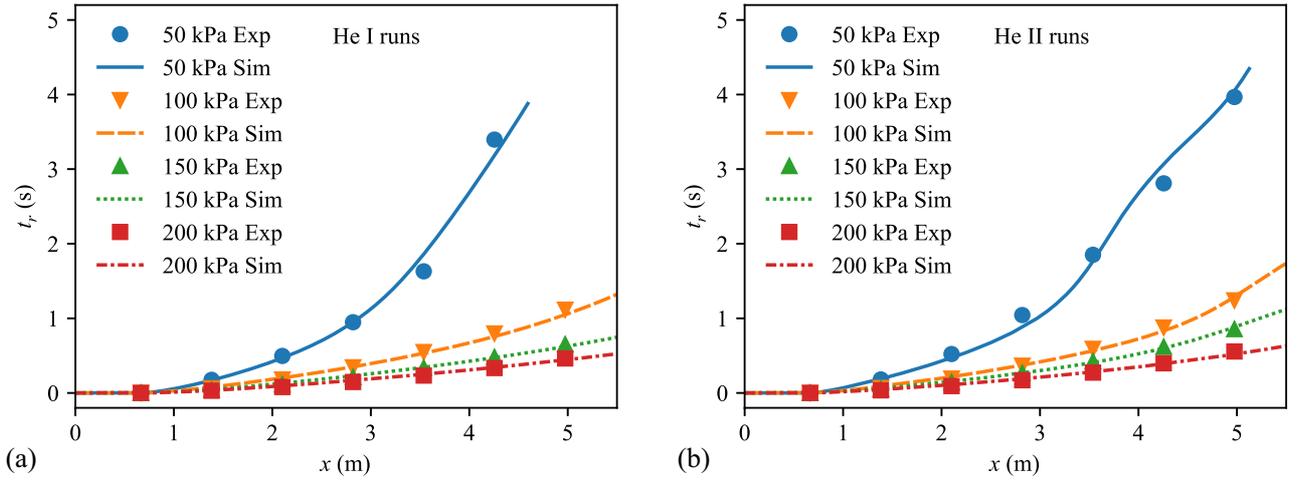}
	\caption{(a) Comparison of the simulated and measured arrival time $t_r(x)$ of the tube-wall temperature versus position $x$ along the tube cooled by He I at a bath temperature of 4.2~K. (b) Comparison of similar simulation and measurement results for the He II runs with a bath temperature of 1.9~K.
		\label{fig:fig3}}
\end{figure*}
The parameter $q_He$ in Eq.~\ref{eq:wall} represents the instantaneous heat flux from the outer surface of the tube to the LHe (He I or He II) bath. Since the temperature relaxation in LHe towards a steady state occurs much faster compared to the slow gas propagation~\cite{van_sciver_helium_2012,Bao-2021-PRB,Sanavandi-2022-PRB}, it is feasible to adopt the correlations developed for steady-state heat transfer between solid surfaces and LHe to model $q_He$ for simplicity~\cite{smith-1969_Cryogenicsa}. The physical properties of GN$_2$, copper, and stainless steel used in our simulations are extracted from existing literature~\cite{stephan1987-Physical-chemical-ref, flynn2004cryogenic, lemmon2007nist, arp2005-INC}. Fig.~\ref{fig:fig2} shows some representative correlation curves for $q_He$. Depending on the temperature difference $\Delta T=T_w-T_b$ between the tube wall and the LHe bath, there are three regimes of $q_He$. In the case of He I, a nucleate boiling regime exists for small $\Delta T$. When $q_He$ reaches a peak heat flux value $q^*$, a transition regime emerges, where $\Delta T$ increases rapidly towards the final film boiling regime. We adopt the Breen-Westwater correlation to evaluate $q^*$~\cite{garceau-2021_InternationalJournalofHeatandMassTransfer,breen1962-chem-eng}. For He II, instead of nucleate boiling, a Kapitza heat transfer regime exists below the peak heat flux $q^*$, whose value can be calculated using an integral correlation that involves both the He II bath temperature $T_b$ and the hydrostatic head pressure (hence the immersion depth $H$)~\cite{van_sciver_helium_2012}. We adopt a two-step first order Godunov-type finite-difference scheme~\cite{danaila2007introduction, sod1978-Journal_computational_physics} to solve the above equations. More detailed information on the assumptions and justifications underlying our 1D theoretical models can be found in our earlier papers~\cite{bao-2020_InternationalJournalofHeatandMassTransfer, garceau-2021_InternationalJournalofHeatandMassTransfer}.

In order to achieve the best agreement between the simulated and observed gas dynamics, we treated the coefficient $B_w$ in the Breen-Westwater correlation \cite{breen1962-chem-eng} of $q^*$ for He I and the parameter $\Psi$ in the integral correlation of $q^*$ for He II as tuning parameters~\cite{bao-2020_InternationalJournalofHeatandMassTransfer, garceau-2021_InternationalJournalofHeatandMassTransfer}. We found that observations in the He I experiments at various inlet mass flow rates can be nicely reproduced with the optimal $B_w=0.021$~W/(cm$^2\cdot$ K$^{5/4}$). For the He II experiments, due to the complication that the immersion depth changes along the experimental tube, the optimal $\Psi$ value appears to depend on the inlet mass flow rate and varies in the range of 0.4 to 2~\cite{garceau-2021_InternationalJournalofHeatandMassTransfer}. To show the effectiveness of these fine-tuned models, we present in Fig.~\ref{fig:fig3} the simulated GN$_{2}$ arrival time $t_r(x)$ under different inlet gas pressures in both He I and He II cooled tube together with relevant experimental data. This arrival time $t_r(x)$ is defined as the time moment when the wall temperature $T_w(x)$ rises sharply from the bath temperature (i.e., 4.2~K for He I and 1.9~K for He II) to above a threshold temperature of 4.7~K~\cite{garceau-2021_InternationalJournalofHeatandMassTransfer}. The tube inner diameter $D_1$ in these simulations is set to one inch to match the experimental setup. The comparison clearly demonstrates that our model simulations effectively reproduce the gas propagation inside the tube for both the He I runs and the He II runs.

\begin{figure}[!tb]
	\centering
	\includegraphics[width=0.9\columnwidth]{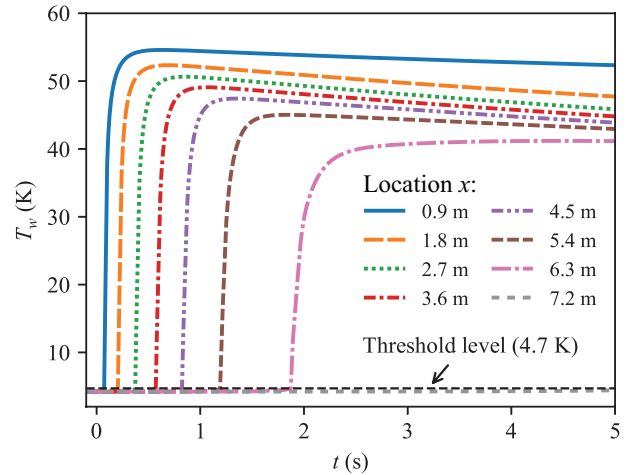}
	\caption{Simulated wall temperature curves $T_w(t)$ at different locations along the copper tube cooled by He I at 4.2~K. $D_1=2$ cm and $\omega=60$ kg/(m$^{2}\cdot$s) are adopted in the model simulation.
		\label{fig:fig4}}
\end{figure}

\section{\label{sec:result}Freeze-range calculation}
\begin{figure}[!tb]
	\centering
	\includegraphics[width=0.9\columnwidth]{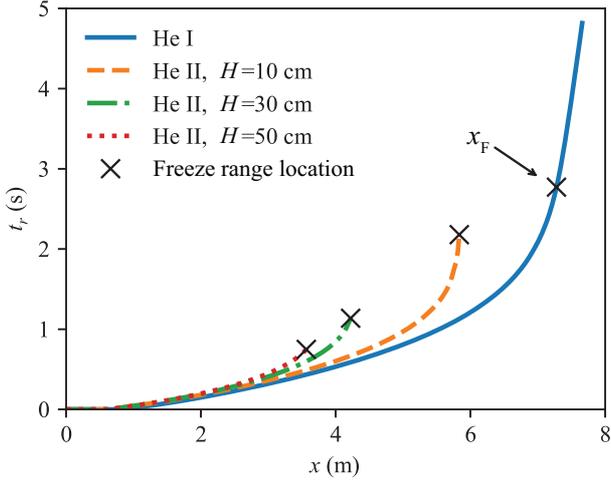}
	\caption{Arrival time $t_r(x)$ as a function of location $x$ along the copper tube cooled by He I at 4.2~K and by He II at 1.9~K with different immersion depth $H$. $D_1=2$ cm and $\omega=60$ kg/(m$^{2}\cdot$s) are adopted in these simulations.
		\label{fig:fig5}}
\end{figure}

\begin{figure}[!tb]
	\centering
	\includegraphics[width=0.9\columnwidth]{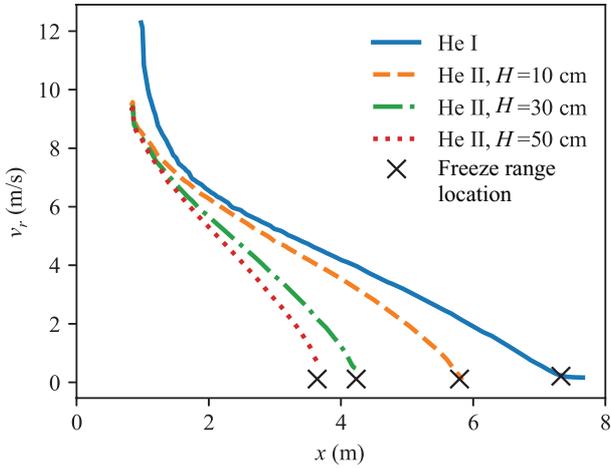}
	\caption{GN$_2$ front propagation velocity $v_r(x)$ as a function of location $x$ along the copper tube cooled by He I at 4.2~K and by He II at 1.9~K with different immersion depth $H$. $D_1=2$ cm and $\omega=60$ kg/(m$^{2}\cdot$s) are adopted in these simulations.
		\label{fig:fig6}}
\end{figure}

In our numerical studies, we have observed an intriguing phenomenon: the propagation of the GN$_2$ appears to nearly stop beyond a certain distance from the entrance of the tube. To illustrate this phenomenon, we show in Fig.~\ref{fig:fig4} the simulated wall temperature curves $T_w(x,t)$ at various locations along a He I cooled tube. These locations are equally spaced by 0.9 m. In this simulation, the tube has an inner diameter of $D_1=2$ cm with a wall thickness of 1.25~mm, and the mass flux of GN$_2$ at the entrance is set to $\omega=\dot m/(\pi D_1^2/4)=$60 kg/(m$^2\cdot$s). As the GN$_2$ front reaches each probed location, there is a sudden rise in the tube wall temperature. To quantify the gas propagation, we extract the arrival time $t_r(x)$ from these wall temperature curves and plot it in Fig.~\ref{fig:fig5}. It is evident that as we move further along the tube (increasing $x$), the arrival time of the gas front increases rapidly. This indicates a deceleration in the gas propagation. Remarkably, as shown in Fig.~\ref{fig:fig4}, the wall temperature at $x=7.2$~m hardly shows any significant increase despite the continuous supply of GN$_2$ at the entrance. This observation strongly suggests that the propagation of GN$_2$ must come to a halt before reaching $x=7.2$ m. To determine the exact location at which the propagation speed diminishes, we calculate the gas-front propagation velocity $v_r=dx/dt_r$ and presented it as a function of $x$ in Fig.~\ref{fig:fig6}. Our analysis reveals a dramatic drop in $v_r$ from over 12~m/s to approximately 7~m/s within a distance of about 1 m from the condensation position (i.e., $x=0.57$~m). Subsequently, $v_r$ continues to decrease roughly linearly with increasing $x$ until it eventually settles at a small value of about 0.18~m/s. We define the distance between the condensation location to the start of this settled regime as the freeze range $x_F$. In the specific case considered here, we find $x_F=6.7$~m, which is depicted in both Fig.~\ref{fig:fig5} and Fig.~\ref{fig:fig6}.


\begin{figure}[!tb]
	\centering
	\includegraphics[width=0.9\columnwidth]{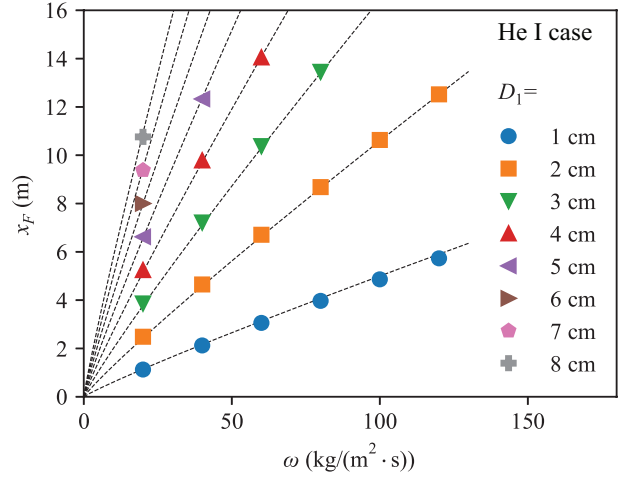}
	\caption{Freeze range $x_F$ for He I cooled tube determined through our model simulations conducted at different mass flux values at the tube entrance $\omega$ and tube inner diameter $D_1$. The dotted lines represent the correlation generated using Eq.~\ref{eq:fitmodel}, which incorporates the optimal parameters listed in Table~\ref{tab:tab1} that yield the best agreement with the simulation data.
		\label{fig:fig7}}
\end{figure}
\begin{figure}[!tb]
	\centering
	\includegraphics[width=0.9\columnwidth]{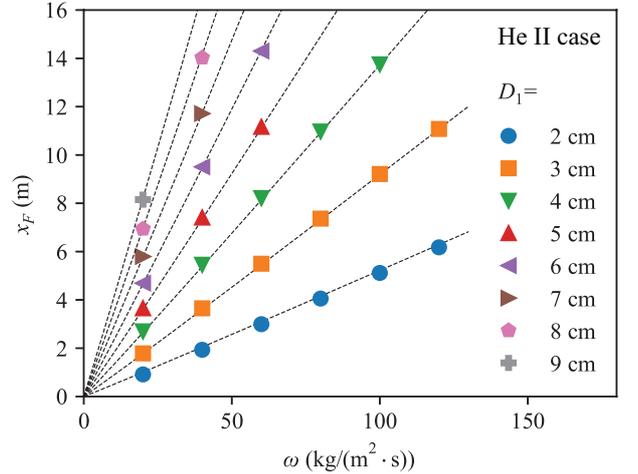}
	\caption{Freeze range $x_F$ for He II cooled tube at different mass flux values at the tube entrance $\omega$ and tube inner diameter $D_1$. The dotted lines represent the correlation generated using Eq.~\ref{eq:fitmodel}, which incorporates the optimal parameters listed in Table~\ref{tab:tab1} that yield the best agreement with the simulation data. The immersion depth $H=50$~cm, and the bath temperature $T_b=1.9$~K.
		\label{fig:fig8}}
\end{figure}

{\centering	
	\begin{table*}[tb]
		\caption{Optimal correlation parameters for evaluating freeze range of propagating GN$_2$ in He I- and He II-cooled copper tube.\label{tab:tab1}}
		\centering
		\begin{tabular}{c c c c}	
			\hline
			Parameter		& Optimal value for He I		& Optimal value for He II (H=50 cm)	& Unit			 		\\
			\hline
			$a$    			& 0.074							& 0.018						& m\textsuperscript{2c-b+1}$\cdot$kg\textsuperscript{-c}$\cdot$s\textsuperscript{c}	\\
			$b$         	& 0.915							& 1.023	 					& 1  		  			\\
			$c$      	 	& 1.084							& 1.395						& 1   		 			\\
			\hline	
		\end{tabular}
	\end{table*}
}

For the He I cooled tube with a fixed wall thickness, the freeze range $x_F$ should depend primarily on the tube inner diameter $D_1$ and the GN$_2$ mass flux $\omega$ at the tube entrance. To explore the dependance of $x_F$ on $D_1$ and $\omega$, we have conducted additional model simulations, varying the values of $D_1$ and $\omega$ within the ranges of 1~cm~$\leq D_1\leq8$~cm and 20 kg/(m$^2\cdot$s)$~\leq\omega\leq120$~kg/(m$^2\cdot$s), respectively. The $x_F$ values obtained from these simulations are presented in Fig.~\ref{fig:fig7}. It is evident that as either $D_1$ or $\omega$ increases, the freeze range $x_F$ also increases. To quantify this relationship, we propose the following simple correlation:
\begin{equation}
	x_F=a\cdot D_1^b\cdot\omega^c,
	\label{eq:fitmodel}
\end{equation}
where $a$, $b$, and $c$ are adjustable parameters. By tuning these parameters, we can effectively reproduce the simulated values of $x_F$ for all combinations of $D_1$ and $\omega$, as illustrated in Fig.~\ref{fig:fig7}. The optimal values of $a$, $b$, and $c$ corresponding to the best fit to the He I data are provided in Table~\ref{tab:tab1}. Notably, both $b$ and $c$ are close to one, suggesting an approximate linear dependance of $x_F$ on $D_1$ and $\omega$.

We have also conducted a detailed analysis of the freeze range for the He II-cooled tube, which presents a more intricate situation. In the case of He II, the heat transfer from the outer surface of the tube to the He II bath is influenced by both the bath temperature $T_b$ and the immersion depth $H$. Consequently, the freeze range $x_F$ is expected to depend not only on the parameters $D_1$ and $\omega$ but also on $T_b$ and $H$. To illustrate this point, we have performed simulations to examine the arrival time $t_r(x)$ and the front propagation velocity $v_r(x)$ of GN$_2$ for a copper tube cooled by He II at various immersion depths $H$, assuming $T_b=1.9$ K. The results are depicted in Fig.~\ref{fig:fig5} and Fig.~\ref{fig:fig6}. The tube's geometry and the mass flux $\omega$ remain unchanged from the He I-cooled tube case. These figures clearly demonstrate that as the immersion depth $H$ increases, the propagation of the GN$_2$ front slows down at a faster rate. At the freeze range $x_F$, denoted by the crosses in the figures, the front propagation velocity $v_r$ drops down to nearly zero. The value of $x_F$ decreases as $H$ increases, which is primarily caused by the enhanced He II peak heat flux $q^*$, as shown in Fig.~\ref{fig:fig2}. Considering the typical operating conditions of SRF cavities, we assume a simplified scenario with fixed $H=50$ cm and $T_b=1.9$ K, and conduct additional simulations to explore the dependence of $x_F$ on $D_1$ and $\omega$. The results of these simulations are presented in Fig.~\ref{fig:fig8}. Once again, we observe that the freeze range $x_F$ increases when either $D_1$ or $\omega$ increases. Interestingly, the correlation expressed by Eq.~\ref{eq:fitmodel} can still provide a fairly good agreement with the simulation data, when the correlation parameters are tuned to the optimal values as listed in Table~\ref{tab:tab1}. These optimal correlation parameters suggest that the freeze range $x_F$ still exhibits a linear dependence on $D_1$, although the dependence on $\omega$ is a bit stronger than a simple linear fashion.

\section{\label{sec:discussion}Discussion}
\begin{figure}[!tb]
	\centering
	\includegraphics[width=0.9\columnwidth]{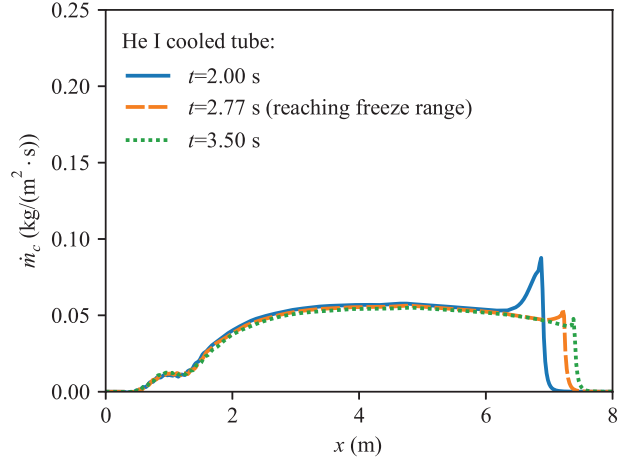}
	\caption{Nitrogen mass deposition rate $\dot m_c(x)$ per unit area on the inner surface of the tube cooled by He I at different propagation times. $D_1=2$ cm and $\omega=60$ kg/(m$^{2}\cdot$s).
		\label{fig:fig9}}
\end{figure}

The approximate linear dependence of $x_F$ on $D_1$ and $\omega$, particularly in the case of a He I-cooled tube, suggests that the nitrogen mass deposition rate $\dot m_c$ per unit area on the tube inner surface should remain relatively constant regardless of the location traversed by the GN$_2$ front. This constancy arises from the fact that once the GN$_2$ front reaches the freeze range, the mass conservation law dictates that the rate at which nitrogen mass is supplied at the tube entrance, i.e., $\dot{m}=(\pi D_1^2/4)\cdot\omega$, must be balanced by the total rate of nitrogen mass deposited on the inner surface of the tube, approximated as $\dot{m}\simeq(x_F\cdot\pi D_1)\cdot\dot m_c$. This balance yields the relationship $x_F\simeq\frac{1}{4\dot m_c}\cdot D_1\cdot\omega$. Consequently, when $\dot m_c$ remains relatively constant, $x_F$ can exhibit a linear dependence on both $D_1$ and $\omega$. To visualize this behavior, Fig.~\ref{fig:fig9} displays the calculated $\dot m_c$ for the He I-cooled tube case as presented in Fig.~\ref{fig:fig4}. Notably, in regions not too close to the condensation location, $\dot m_c$ remains relatively flat, particularly after the GN$_2$ front reaches the freeze range.

However, the constancy of $\dot m_c$ independent of position $x$, $\omega$, and $D_1$ may seem counterintuitive, given that according to Eq.~\ref{eq:mc}, $\dot m_c$ depends on the GN$_2$ pressure $P$, temperature $T_g$, and the surface temperature of the frost layer $T_s$. These parameters exhibit significant variations within the freeze range and are also influenced by the mass flux $\omega$ at the entrance. To shed light on the underlying physical mechanism, we examine heat transfer in the radial direction through the wall into the LHe bath. Upon the arrival of the GN$_2$ front, the temperature $T_w(x)$ of the wall at each location increases to approximately 40~K to 55~K and changes slowly over time, as shown in Fig.~\ref{fig:fig4}. Due to the thinness of the frost layer, $T_s(x)$ remains close to $T_w(x)$ at all times~\cite{bao-2020_InternationalJournalofHeatandMassTransfer}. Consequently, we can disregard the time derivative terms in Eq.~(\ref{eq:frost}) and Eq.~(\ref{eq:wall}) and derive:
\begin{equation}
	\dot{m}_c[\frac{1}{2}v^2+\hat{h}_g-\hat{h}_s]+Nu\cdot k_g\frac{(T_g-T_s)}{D_1}=q_{He}\frac{D_2}{D_1}\simeq q_{He}.
	\label{eq:heatbalance}
\end{equation}
The last term on the right-hand side of Eq.~(\ref{eq:wall}), which represents heat conduction along the tube, is negligible compared to other terms and is therefore omitted in the derivation of the above equation. Essentially, Eq.~(\ref{eq:heatbalance}) describes a local balance between the heat flux deposited on the tube wall and the heat flux into the LHe bath.

\begin{figure}[!tb]
	\centering
	\includegraphics[width=0.9\columnwidth]{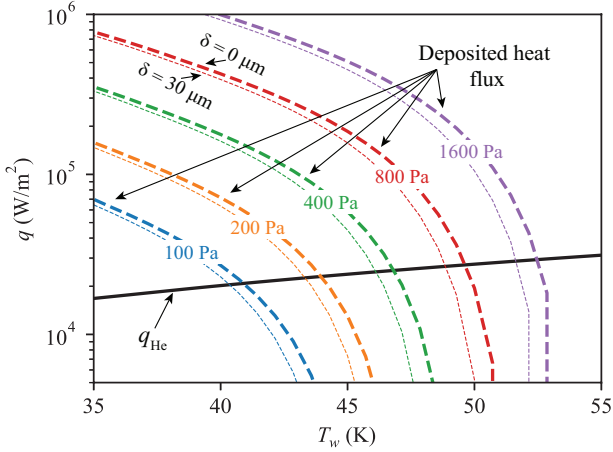}
	\caption{Variation of the heat flux deposited on the tube wall (i.e., the left-hand side of Eq.~(\ref{eq:heatbalance})) and the heat flux into the He I bath (i.e., $q_{He}$) at different wall temperature $T_w$. The dotted curves represent the deposited heat flux when there is a frost layer of $\delta=30$~$\mu$m on the tube inner surface.
		\label{fig:fig10}}
\end{figure}

When the bath temperature $T_b$ remains constant, the heat flux $q_{He}$ into the bath depends solely on the tube wall temperature $T_w$ (refer to Fig.~\ref{fig:fig2}), which increases with increasing $T_w$ as illustrated in Fig.~\ref{fig:fig10}. On the other hand, as $T_w$ (and thus $T_s$) rises, the mass deposition rate $\dot{m}_c$ decreases, resulting in a reduced rate of heat deposition on the tube wall. For given GN$_2$ state parameters $P$ and $T_g$, Eq.~(\ref{eq:heatbalance}) can be balanced at a specific $T_w$. In Fig.~\ref{fig:fig10}, we present curves showing the variation of the deposited heat flux (i.e., the left-hand side of Eq.~(\ref{eq:heatbalance})) as a function of $T_w$ at different $P$. In these calculations, we set $T_g=100$~K and assume $T_s\simeq T_w$. Evidently, the equilibrium $T_w$ increases with higher $P$, consistent with the trend observed in Fig.~\ref{fig:fig4}, where a higher $T_w$ is observed closer to the tube entrance where $P$ is larger. Remarkably, when $P$ increases from 100~Pa to 1600~Pa, $T_w$ only rises by about 10~K, resulting in less than a 25\% increase in $q_{He}$. Similar results are obtained at other $T_g$ values, and the effect due to the difference between $T_s$ and $T_w$ caused by a finite frost layer on the tube inner surface is also insignificant (see Fig.~\ref{fig:fig10}). The insensitivity of $q_{He}$ to the state parameters of the GN$_2$ implies that it should remain roughly constant independent of $x$ and $\omega$. In the region near the condensation location (i.e., 0.57~m~$\leq x\leq2$~m), the convective heat transfer term on the left-hand side of Eq.~(\ref{eq:heatbalance}) is significant. Consequently, the value of $\dot{m}_c$ in this region is relatively low, as evident in Fig.~\ref{fig:fig9}. However, as we move away from this region, Eq.~(\ref{eq:heatbalance}) can be simplified to $\dot{m}_c\simeq q_{He}/(\hat{h}_g-\hat{h}_s)$. As a result, the stability of the heat flux $q_{He}$ leads to a relatively stable value for $\dot{m}_c$, which in turn leads to the approximate linear dependance of $x_F$ on $D_1$ and $\omega$.

\section{\label{sec:sum}Conclusion}
We have conducted a comprehensive numerical investigation into the freeze range phenomenon that occurs during the propagation of GN$_2$ in a copper tube cooled by LHe. Our study reveals that the freeze range $x_F$, within which the nitrogen condensation occurs primarily, exhibits an increase with both the tube diameter and the mass flux at the tube entrance. Through a careful analysis of $x_F$ values obtained for various tube diameters and mass fluxes at the entrance, we have developed a simple correlation that demonstrates good agreement with our numerical simulation results. This correlation suggests that the rate of nitrogen condensation on the inner surface of the tube remains relatively constant and is not significantly affected by changes in tube diameter, positions along the tube, or the mass flux at the entrance. Further analysis has uncovered a local balance between the heat flux deposited on the tube wall and the heat flux transferred into the LHe bath after the GN$_2$ front reaches the freeze range. Notably, the heat flux into the LHe bath exhibits high insensitivity to variations in GN$_2$ state parameters, resulting in the nearly constant condensation rate. Our findings, particularly the establishment of the simple correlation, hold practical value for accelerator engineers who are involved in the development of protocols for managing frost-layer contamination in accelerator beamline tubes. These insights can assist in formulating effective strategies to control and mitigate the impact of frost-layer formation in such systems.

\section*{Acknowledgments}
This work is supported by U.S. Department of Energy under Grant No. DE-SC0020113. The experiment was conducted at the National High Magnetic Field Laboratory, which is supported by National Science Foundation Cooperative Agreement No. DMR-2128556 and the State of Florida.







\bibliographystyle{elsarticle-num-names}


\begin{thebibliography}{32}
\expandafter\ifx\csname natexlab\endcsname\relax\def\natexlab#1{#1}\fi
\providecommand{\url}[1]{\texttt{#1}}
\providecommand{\href}[2]{#2}
\providecommand{\path}[1]{#1}
\providecommand{\DOIprefix}{doi:}
\providecommand{\ArXivprefix}{arXiv:}
\providecommand{\URLprefix}{URL: }
\providecommand{\Pubmedprefix}{pmid:}
\providecommand{\doi}[1]{\href{http://dx.doi.org/#1}{\path{#1}}}
\providecommand{\Pubmed}[1]{\href{pmid:#1}{\path{#1}}}
\providecommand{\bibinfo}[2]{#2}
\ifx\xfnm\relax \def\xfnm[#1]{\unskip,\space#1}\fi
\bibitem[{Padamsee(2009)}]{Padamsee-2009-RFSuperCon}
\bibinfo{author}{H.~Padamsee}, \bibinfo{title}{{{RF Superconductivity}}:
  {{Science}}, {{Technology}}, and {{Applications}}}, \bibinfo{publisher}{{John
  Wiley \& Sons}}, \bibinfo{year}{2009}.
\bibitem[{Pagani et~al.(2005)Pagani, Pierini et~al.}]{pagani-2005-SRF05}
\bibinfo{author}{C.~Pagani}, \bibinfo{author}{P.~Pierini}, et~al.,
\newblock \bibinfo{title}{Cryomodule design, assembly and alignment},
\newblock \bibinfo{journal}{SRF05, Ithaca NY, USA}  (\bibinfo{year}{2005}).
\bibitem[{Wiseman et~al.(1994)Wiseman, Crawford, Drury, Jordan, Preble,
  Saulter, and Schneider}]{wiseman1994loss}
\bibinfo{author}{M.~Wiseman}, \bibinfo{author}{K.~Crawford},
  \bibinfo{author}{M.~Drury}, \bibinfo{author}{K.~Jordan},
  \bibinfo{author}{J.~Preble}, \bibinfo{author}{Q.~Saulter},
  \bibinfo{author}{W.~Schneider},
\newblock \bibinfo{title}{Loss of cavity vacuum experiment at {CEBAF}},
\newblock \bibinfo{journal}{Advances in cryogenic engineering}
  (\bibinfo{year}{1994}) \bibinfo{pages}{997--1003}.
  \DOIprefix\doi{10.1007/978-1-4615-2522-6\_121}.
\bibitem[{Seidel et~al.(2002)Seidel, Trines, and Zapfe}]{seidel2002failure}
\bibinfo{author}{M.~Seidel}, \bibinfo{author}{D.~Trines},
  \bibinfo{author}{K.~Zapfe}, \bibinfo{title}{Failure Analysis of the Beam
  Vacuum in the Superconducting Cavities of the {{TESLA}} Main Linear
  Accelerator}, \bibinfo{type}{{TESLA-Report}} \bibinfo{number}{2002-06},
  {Deutsches Elektronen Synchrotron DESY}, \bibinfo{address}{{Hamburg,
  Germany}}, \bibinfo{year}{2002}.
\bibitem[{Ady(2014)}]{ady2014-CERN}
\bibinfo{author}{M.~Ady}, \bibinfo{title}{Measurement Campaign of 21-25
  {{July}} 2014 for Evaluation of the {{HIE-ISOLDE}} Inrush Protection System},
  \bibinfo{type}{{{CERN}} Technical Note} \bibinfo{number}{TE-VSC-1414574},
  {Technology Department, CERN}, \bibinfo{address}{{Switzerland}},
  \bibinfo{year}{2014}.
\bibitem[{Boeckmann et~al.(2008)Boeckmann, Hoppe, Jensch, Lange, Maschmann,
  Petersen, and Schnautz}]{boeckmann2008-ICE}
\bibinfo{author}{T.~Boeckmann}, \bibinfo{author}{D.~Hoppe},
  \bibinfo{author}{K.~Jensch}, \bibinfo{author}{R.~Lange},
  \bibinfo{author}{W.~Maschmann}, \bibinfo{author}{B.~Petersen},
  \bibinfo{author}{T.~Schnautz},
\newblock \bibinfo{title}{Experimental tests of fault conditions during the
  cryogenic operation of a {{XFEL}} prototype cryomodule},
\newblock in: \bibinfo{booktitle}{Proceedings of International Cryogenic
  Engineering Conference}, volume~\bibinfo{volume}{22},
  \bibinfo{address}{{Seoul, Korea}}, \bibinfo{year}{2008}, pp.
  \bibinfo{pages}{723--728}.
\bibitem[{Dalesandro et~al.(2012)Dalesandro, Theilacker, and
  Van~Sciver}]{dalesandro2012-AIP}
\bibinfo{author}{A.~A. Dalesandro}, \bibinfo{author}{J.~Theilacker},
  \bibinfo{author}{S.~W. Van~Sciver},
\newblock \bibinfo{title}{Experiment for transient effects of sudden
  catastrophic loss of vacuum on a scaled superconducting radio frequency
  cryomodule},
\newblock in: \bibinfo{booktitle}{AIP Conference Proceedings}, volume
  \bibinfo{volume}{1434}, \bibinfo{organization}{American Institute of
  Physics}, \bibinfo{year}{2012}, pp. \bibinfo{pages}{1567--1574}.
  \DOIprefix\doi{10.1063/1.4707087}.
\bibitem[{Dalesandro(2023)}]{dalesandro2023}
\bibinfo{author}{A.~A. Dalesandro}, \bibinfo{title}{Internal review on the
  state of loss of vacuum research on helium cryogenic systems},
  \bibinfo{year}{2023}. \DOIprefix\doi{10.48550/arXiv.2303.15309}.
  \href{http://arxiv.org/abs/2303.15309}{{\tt arXiv:2303.15309}}.
\bibitem[{Boeckmann et~al.(2008)Boeckmann, Hoppe, Jensch, Lange, Maschmann,
  Petersen, and Schnautz}]{boeckmann-2008_Int.Cryog.Eng.Conf.}
\bibinfo{author}{T.~Boeckmann}, \bibinfo{author}{D.~Hoppe},
  \bibinfo{author}{K.~Jensch}, \bibinfo{author}{R.~Lange},
  \bibinfo{author}{W.~Maschmann}, \bibinfo{author}{B.~Petersen},
  \bibinfo{author}{T.~Schnautz},
\newblock \bibinfo{title}{Experimental tests of fault conditions during the
  cryogenic operation of a {{XFEL}} prototype cryomodule.},
\newblock in: \bibinfo{booktitle}{International {{Cryogenic Engineering
  Conference}}}, volume~\bibinfo{volume}{22}, \bibinfo{address}{{Seoul,
  Korea}}, \bibinfo{year}{2008}, pp. \bibinfo{pages}{723--728}.
\bibitem[{Ady et~al.(2014)Ady, Hermann, Kersevan, Vandoni, and
  Ziemianski}]{Ady-2014-5th-IPAC-Proc.}
\bibinfo{author}{M.~Ady}, \bibinfo{author}{M.~Hermann},
  \bibinfo{author}{R.~Kersevan}, \bibinfo{author}{G.~Vandoni},
  \bibinfo{author}{D.~Ziemianski},
\newblock \bibinfo{title}{Leak propagation dynamics for the {{HIE}}-{{ISOLDE}}
  superconducting linac},
\newblock in: \bibinfo{editor}{P.-J.-G. Christine~(Ed.)},
  \bibinfo{editor}{A.~Gianluigi~(Ed.)}, \bibinfo{editor}{M.~Peter~(Ed.)},
  \bibinfo{editor}{R.~S. Volker(Ed.)} (Eds.), \bibinfo{booktitle}{Proceedings
  of the 5th {{Int}}. {{Particle Accelerator Conf}}.},
  \bibinfo{address}{{Dresden, Germany}}, \bibinfo{year}{2014}, pp.
  \bibinfo{pages}{2351--2353}.
  \DOIprefix\doi{10.18429/JACoW-IPAC2014-WEPME039}.
\bibitem[{Dhuley and
  Van~Sciver(2016{\natexlab{a}})}]{dhuley-2016_Int.J.HeatMassTransf.}
\bibinfo{author}{R.~C. Dhuley}, \bibinfo{author}{S.~W. Van~Sciver},
\newblock \bibinfo{title}{Propagation of nitrogen gas in a liquid helium cooled
  vacuum tube following sudden vacuum loss \textendash{} {{Part I}}:
  {{Experimental}} investigations and analytical modeling},
\newblock \bibinfo{journal}{International Journal of Heat and Mass Transfer}
  \bibinfo{volume}{96} (\bibinfo{year}{2016}{\natexlab{a}})
  \bibinfo{pages}{573--581}.
  \DOIprefix\doi{10.1016/j.ijheatmasstransfer.2016.01.077}.
\bibitem[{Dhuley and
  Van~Sciver(2016{\natexlab{b}})}]{dhuley-2016_Int.Cryog.Eng.Conf.2}
\bibinfo{author}{R.~C. Dhuley}, \bibinfo{author}{S.~W. Van~Sciver},
\newblock \bibinfo{title}{Propagation of nitrogen gas in a liquid helium cooled
  vacuum tube following sudden vacuum loss \textendash{} {{Part II}}:
  {{Analysis}} of the propagation speed},
\newblock \bibinfo{journal}{International Journal of Heat and Mass Transfer}
  \bibinfo{volume}{98} (\bibinfo{year}{2016}{\natexlab{b}})
  \bibinfo{pages}{728--737}. \DOIprefix\doi{10.1016/j
  .ijheatmasstransfer.2016.03.077}.
\bibitem[{Garceau et~al.(2019{\natexlab{a}})Garceau, Bao, Guo, and
  Van~Sciver}]{garceau-2019_Cryogenics}
\bibinfo{author}{N.~Garceau}, \bibinfo{author}{S.~R. Bao},
  \bibinfo{author}{W.~Guo}, \bibinfo{author}{S.~W. Van~Sciver},
\newblock \bibinfo{title}{The design and testing of a liquid helium cooled tube
  system for simulating sudden vacuum loss in particle accelerators},
\newblock \bibinfo{journal}{Cryogenics} \bibinfo{volume}{100}
  (\bibinfo{year}{2019}{\natexlab{a}}) \bibinfo{pages}{92--96}.
  \DOIprefix\doi{10.1016/j.cryogenics.2019.04.012}.
\bibitem[{Garceau et~al.(2019{\natexlab{b}})Garceau, Bao, and
  Guo}]{garceau-2019_InternationalJournalofHeatandMassTransfer}
\bibinfo{author}{N.~Garceau}, \bibinfo{author}{S.~R. Bao},
  \bibinfo{author}{W.~Guo},
\newblock \bibinfo{title}{Heat and mass transfer during a sudden loss of vacuum
  in a liquid helium cooled tube \textendash{} {{Part I}}: {{Interpretation}}
  of experimental observations},
\newblock \bibinfo{journal}{International Journal of Heat and Mass Transfer}
  \bibinfo{volume}{129} (\bibinfo{year}{2019}{\natexlab{b}})
  \bibinfo{pages}{1144--1150}. \DOIprefix\doi{10.1016/j
  .ijheatmasstransfer.2018.10.053}.
\bibitem[{Bao et~al.(2020)Bao, Garceau, and
  Guo}]{bao-2020_InternationalJournalofHeatandMassTransfer}
\bibinfo{author}{S.~R. Bao}, \bibinfo{author}{N.~Garceau},
  \bibinfo{author}{W.~Guo},
\newblock \bibinfo{title}{Heat and mass transfer during a sudden loss of vacuum
  in a liquid helium cooled tube \textendash{} {{Part II}}: {{Theoretical}}
  modeling},
\newblock \bibinfo{journal}{International Journal of Heat and Mass Transfer}
  \bibinfo{volume}{146} (\bibinfo{year}{2020}) \bibinfo{pages}{118883}.
  \DOIprefix\doi{10.1016/j .ijheatmasstransfer.2019.118883}.
\bibitem[{Garceau et~al.(2021)Garceau, Bao, and
  Guo}]{garceau-2021_InternationalJournalofHeatandMassTransfer}
\bibinfo{author}{N.~Garceau}, \bibinfo{author}{S.~R. Bao},
  \bibinfo{author}{W.~Guo},
\newblock \bibinfo{title}{Heat and mass transfer during a sudden loss of vacuum
  in a liquid helium cooled tube - {{Part III}}: {{Heat}} deposition in {{He
  II}}},
\newblock \bibinfo{journal}{International Journal of Heat and Mass Transfer}
  \bibinfo{volume}{181} (\bibinfo{year}{2021}) \bibinfo{pages}{121885}.
  \DOIprefix\doi{10.1016/j .ijheatmasstransfer.2021.121885}.
\bibitem[{Incropera et~al.(2007)Incropera, DeWitt, Bergman, Lavine
  et~al.}]{incropera-2007_book}
\bibinfo{author}{F.~P. Incropera}, \bibinfo{author}{D.~P. DeWitt},
  \bibinfo{author}{T.~L. Bergman}, \bibinfo{author}{A.~S. Lavine}, et~al.,
  \bibinfo{title}{Fundamentals of Heat and Mass Transfer},
  volume~\bibinfo{volume}{6}, \bibinfo{edition}{6} ed.,
  \bibinfo{publisher}{{John Wiley \& Sons}}, \bibinfo{address}{{Hoboken, NJ}},
  \bibinfo{year}{2007}.
\bibitem[{Collier and Thome(1994)}]{collier1994convective}
\bibinfo{author}{J.~G. Collier}, \bibinfo{author}{J.~R. Thome},
  \bibinfo{title}{Convective boiling and condensation},
  \bibinfo{publisher}{Clarendon Press}, \bibinfo{year}{1994}.
\bibitem[{Persad and Ward(2016)}]{persad-2016_Chem.Rev.}
\bibinfo{author}{A.~H. Persad}, \bibinfo{author}{C.~A. Ward},
\newblock \bibinfo{title}{Expressions for the evaporation and condensation
  coefficients in the {{Hertz}}-{{Knudsen}} relation},
\newblock \bibinfo{journal}{Chemical reviews} \bibinfo{volume}{116}
  (\bibinfo{year}{2016}) \bibinfo{pages}{7727--7767}.
  \DOIprefix\doi{10.1021/acs.chemrev.5b00511}.
\bibitem[{Stephan et~al.(1987)Stephan, Krauss, and
  Laesecke}]{stephan1987-Physical-chemical-ref}
\bibinfo{author}{K.~Stephan}, \bibinfo{author}{R.~Krauss},
  \bibinfo{author}{A.~Laesecke},
\newblock \bibinfo{title}{Viscosity and thermal conductivity of nitrogen for a
  wide range of fluid states},
\newblock \bibinfo{journal}{Journal of physical and chemical reference data}
  \bibinfo{volume}{16} (\bibinfo{year}{1987}) \bibinfo{pages}{993--1023}.
  \DOIprefix\doi{10.1063/1.555798}.
\bibitem[{Scott(1976)}]{scott1976-Physics_report}
\bibinfo{author}{T.~A. Scott},
\newblock \bibinfo{title}{Solid and liquid nitrogen},
\newblock \bibinfo{journal}{Physics Reports} \bibinfo{volume}{27}
  (\bibinfo{year}{1976}) \bibinfo{pages}{89--157}.
\bibitem[{Cook and Davey(1976)}]{cook-1976_Cryogenicsa}
\bibinfo{author}{T.~Cook}, \bibinfo{author}{G.~Davey},
\newblock \bibinfo{title}{The density and thermal conductivity of solid
  nitrogen and carbon dioxide},
\newblock \bibinfo{journal}{Cryogenics} \bibinfo{volume}{16}
  (\bibinfo{year}{1976}) \bibinfo{pages}{363--369}.
  \DOIprefix\doi{10.1016/0011-2275(76)90217-4}.
\bibitem[{Van~Sciver(2012)}]{van_sciver_helium_2012}
\bibinfo{author}{S.~W. Van~Sciver}, \bibinfo{title}{Helium Cryogenics},
  International cryogenics monograph series, \bibinfo{edition}{2} ed.,
  \bibinfo{publisher}{Springer}, \bibinfo{address}{New York, USA},
  \bibinfo{year}{2012}.
\bibitem[{Bao and Guo(2021)}]{Bao-2021-PRB}
\bibinfo{author}{S.~R. Bao}, \bibinfo{author}{W.~Guo},
\newblock \bibinfo{title}{Transient heat transfer of superfluid
  $^{4}\mathrm{He}$ in nonhomogeneous geometries: Second sound, rarefaction,
  and thermal layer},
\newblock \bibinfo{journal}{Phys. Rev. B} \bibinfo{volume}{103}
  (\bibinfo{year}{2021}) \bibinfo{pages}{134510}.
  \DOIprefix\doi{10.1103/PhysRevB.103.134510}.
\bibitem[{Sanavandi et~al.(2022)Sanavandi, Hulse, Bao, Tang, and
  Guo}]{Sanavandi-2022-PRB}
\bibinfo{author}{H.~Sanavandi}, \bibinfo{author}{M.~Hulse},
  \bibinfo{author}{S.~R. Bao}, \bibinfo{author}{Y.~Tang},
  \bibinfo{author}{W.~Guo},
\newblock \bibinfo{title}{Boiling and cavitation caused by transient heat
  transfer in superfluid helium-4},
\newblock \bibinfo{journal}{Phys. Rev. B} \bibinfo{volume}{106}
  (\bibinfo{year}{2022}) \bibinfo{pages}{054501}.
  \DOIprefix\doi{10.1103/PhysRevB.106.054501}.
\bibitem[{Smith(1969)}]{smith-1969_Cryogenicsa}
\bibinfo{author}{R.~Smith},
\newblock \bibinfo{title}{Review of heat transfer to helium {{I}}},
\newblock \bibinfo{journal}{Cryogenics} \bibinfo{volume}{9}
  (\bibinfo{year}{1969}) \bibinfo{pages}{11--19}.
  \DOIprefix\doi{10.1016/0011-2275(69)90251-3}.
\bibitem[{Flynn(2004)}]{flynn2004cryogenic}
\bibinfo{author}{T.~Flynn}, \bibinfo{title}{Cryogenic engineering},
  \bibinfo{publisher}{CRC Press}, \bibinfo{year}{2004}.
\bibitem[{Lemmon et~al.(2007)Lemmon, Huber, and McLinden}]{lemmon2007nist}
\bibinfo{author}{E.~Lemmon}, \bibinfo{author}{M.~L. Huber},
  \bibinfo{author}{M.~O. McLinden}, \bibinfo{title}{{NIST} standard reference
  database 23: reference fluid thermodynamic and transport
  properties-{{REFPROP}}, version 8.0}, \bibinfo{year}{2007}.
\bibitem[{Arp et~al.(2005)Arp, McCarty, and Jeffrey}]{arp2005-INC}
\bibinfo{author}{V.~Arp}, \bibinfo{author}{R.~McCarty},
  \bibinfo{author}{F.~Jeffrey}, \bibinfo{title}{{HEPAK}}, \bibinfo{year}{2005}.
\bibitem[{Breen et~al.(1962)Breen, Westwater et~al.}]{breen1962-chem-eng}
\bibinfo{author}{B.~Breen}, \bibinfo{author}{J.~Westwater}, et~al.,
\newblock \bibinfo{title}{Effect of diameter of horizontal tubes on film
  boiling heat transfer},
\newblock \bibinfo{journal}{Chemical Engineering Progress} \bibinfo{volume}{58}
  (\bibinfo{year}{1962}) \bibinfo{pages}{67--72}.
\bibitem[{Danaila et~al.(2007)Danaila, Joly, Kaber, and
  Postel}]{danaila2007introduction}
\bibinfo{author}{I.~Danaila}, \bibinfo{author}{P.~Joly}, \bibinfo{author}{S.~M.
  Kaber}, \bibinfo{author}{M.~Postel}, \bibinfo{title}{An introduction to
  scientific computing: Twelve computational projects solved with MATLAB},
  \bibinfo{publisher}{Springer}, \bibinfo{year}{2007}.
  \DOIprefix\doi{10.1007/978-0-387-49159-2}.
\bibitem[{Sod(1978)}]{sod1978-Journal_computational_physics}
\bibinfo{author}{G.~A. Sod},
\newblock \bibinfo{title}{A survey of several finite difference methods for
  systems of nonlinear hyperbolic conservation laws},
\newblock \bibinfo{journal}{Journal of computational physics}
  \bibinfo{volume}{27} (\bibinfo{year}{1978}) \bibinfo{pages}{1--31}.
  \DOIprefix\doi{10.1016/0021-9991(78)90023-2}.

\end{thebibliography}

\section*{Nomenclature}


\tabletail{%
	\hline
}

\tablehead{%
	\hline
}

\begin{tabular}[h]{p{1cm}p{4cm}p{2.1cm}}
	\hline
	Variable 			& Description										   & Units\\
	\hline
	$B_w$       		& Coefficient in the film boiling correlation of He I  & W/(cm$^2\cdot$ K$^{5/4}$)   \\
	$C$       	    	& Specific heat                                        & J/(kg$\cdot $K)   \\
	$D_1$           	& Inner diameter of the tube                           & m                 \\
	$D_2$           	& Outer diameter of the tube                           & m                 \\
	$\hat{h}$           & Specific enthalpy                                    & J/kg              \\
	$H$           		& Immersion depth                                      & m	               \\
	$k$             	& Thermal conductivity                                 & W/(m$\cdot $K)    \\
	$\dot m_c$   & Mass deposition rate                                 & kg/(m$^2\cdot $s) \\
	$\dot m$   			& Mass flow rate                                 	   & kg/s			   \\
	$M_g$             	& Gas molar mass                                       & kg/mol            \\
	$Nu$            	& Nusselt number                                       &                   \\
	$P$             	& Nitrogen gas pressure                                & Pa                \\
	$q$           		& Heat flux											   & W/m$^2$           \\
	$q^*$           	& Peak heat flux                                       & W/m$^2$           \\
	$q_{con}$      & Conductive heat flux along the wall                  & W/m$^2$           \\
	$q_{He}$       & Heat flux to the liquid helium bath                  & W/m$^2$           \\
	$R$             	& Ideal gas constant                                   & J/(mol$\cdot $K)  \\
	$t$             	& Time                                                 & s                 \\
	$t_r$             	& Arrival time of gas front                            & s                 \\
	$T$             	& Temperature                                          & K                 \\
	$T_b$        & Helium bath temperature                              & K                 \\
	$T_c$        & Center temperature of the SN$_2$ layer               & K                 \\
	$v$             	& Gas velocity                                         & m/s               \\
	$v_r$             	& Front propagation velocity                           & m/s               \\
	$x$             	& Axial coordinate                                     & m                 \\
	$x_F$        & Freeze range                   	                   & m                 \\
	
						&                                                      &                   \\
	$Greeks$        	&                                                      &                   \\
	$\Gamma$        	& Schrage parameter                                    &                   \\
	$\delta$        	& Thickness of the SN$_2$ layer                        & m                 \\
	$\dot{\delta}$      & Change rate of $\delta$                        	   & m/s               \\
	$\Delta T$        	& Temperature difference                        	   & K                 \\
	$\varepsilon$   	& Specific internal energy                             & J/kg              \\
	$\rho$          	& Density                                              & kg/m$^3$          \\
	$\sigma_c$	& Condensation coefficient                             & 		           \\
	$\sigma_e$	& Evaporation coefficient                              & 		           \\
	$\Psi$          	& Parameter in the integral correlation of $q^*$ for He II                & kg/(m$^2\cdot $s) \\
	$\omega$          	& Inlet mass flux                                      & kg/(m$^2\cdot $s) \\
						&                                                      &                   \\
	$Subscripts$    	&                                                      &                   \\
	$g$          & Bulk gas condition                                   &                   \\
	$s$          & Surface of SN$_2$ layer           				   &                   \\
	$w$          & Copper tube wall                                     &                   \\
	$SN$         & Solid nitrogen                                       &                   \\
	\hline
\end{tabular}

\end{document}